\documentclass[epj]{svjour}
\usepackage{latexsym}
\usepackage{graphics}

\def\la{\langle}
\def\ra{\rangle}
\def\beq{\begin{equation}}
\def\eeq{\end{equation}}
\def\be{\begin{eqnarray}}
\def\ee{\end{eqnarray}}

\def\k2av{\la k_T^2\ra}
\newcommand{\f}[2]{\frac{#1}{#2}}
\newcommand{\dd}{ {\textrm d}}

\begin{document}

\title{Jet Tomography Studies in $AuAu$ Collisions at RHIC Energies}
%
\author{G.G. Barnaf\"oldi\inst{1,}\inst{2} \and P. L\'evai\inst{1} \and 
G. Papp\inst{3} \and G. F\'ai\inst{4} \and M. Gyulassy\inst{5}
%
}                     
%
%
\institute{
RMKI KFKI, P.O. Box 49, Budapest 1525, Hungary \and
Lab. for Information Technology, E\"otv\"os University, P\'azm\'any P. 1/A, Budapest 1117, Hungary \and 
Dept. for Theoretical Physics, E\"otv\"os University, P\'azm\'any P. 1/A, Budapest 1117, Hungary \and 
CNR, Kent State University, Kent, OH 44242, USA \and
Dept. of Physics, Columbia University, 538 W. 120\textsuperscript{th} 
Street, New York, NY 10027, USA } 
%
%
\abstract{
Recent RHIC results on pion production in $AuAu$ collision 
at $\sqrt{s}=130$ and $200$ AGeV display a 
strong suppression effect at high $p_T$. This 
suppression can be connected to final state effects,
namely jet energy loss
induced by the produced dense colored matter. Applying 
our pQCD-based parton model we perform a 
quantitative analysis of the measured suppression 
pattern and determine the opacity
of the produced deconfined matter. 
} 

\maketitle

\section{Introduction}
\label{intro}

The experimental data on high-$p_T$ $\pi^0$ production 
in central $AuAu$ collisions at mid-rapidity at
$\sqrt{s}=130$ and $200$ AGeV have shown a strong suppression compared
to  binary scaled $pp$ data \cite{phenix0203,star0203}. This suppression 
vanishes with increasing centrality and no effect appears in 
peripheral $AuAu$ collisions.  Thus a detailed quantitative analysis of the 
suppression pattern ({\it "jet tomography"}~\cite{tomog}) 
yields information about the 
properties of the produced dense matter and the impact parameter dependence
of the formation of deconfined matter in $AuAu$ collisions.
Recent data on $dAu$ collisions~\cite{phenixdau,stardau} validate our effort:
since no suppression was found at mid-rapidity in  the $dAu$ reaction, 
initial state 
effects (e.g. a strong modification in the internal parton structure of the 
accelerated $Au$ nuclei) can not be responsible for the measured suppression
in $AuAu$ collisions. Thus, induced jet-energy loss in the final state becomes
a strong candidate to explain the missing pion yield.
Jet energy loss can be calculated in a perturbative 
quantum chromodynamics (pQCD) frame~\cite{gw,glv}. 

We investigate jet energy loss in a pQCD improved parton model. 
In Sect.~\ref{sec:1} we introduce the basis of our model, especially a 
phenomenological intrinsic transverse momentum distribution (intrinsic $k_T$) 
for the colliding nucleons, which is necessary
to reach a better agreement between data and calculations 
in $pp$ collisions~\cite{Wong98,Wang01,Yi02,Bp02}. 
For nucleus-nucleus ($AA$) collisions, initial
state effects are considered, e.g.  nuclear multiple 
scattering, saturation in the number of semihard
collisions, and a weak shadowing effect inside the nucleus~\cite{Wang01,Yi02}.
In Sect.~\ref{sec:2} we summarize the description of induced gluon
radiation in thin non-Abelian matter and include the
GLV-description~\cite{glv} of energy loss into the pQCD improved
parton model. This way our model becomes capable to 
extract the opacity values of the produced 
colored matter at different centralities.
In Sect.~\ref{sec:3} we discuss the obtained results.


\section{Initial State Effects in $AuAu$ Collisions}
\label{sec:1}

The invariant cross section of pion production in an $AA'$ collision
can be described in a pQCD-improved parton model 
developed for $pp$ collision and 
extended by a Glauber-type collision geometry and initial 
state nuclear effects for $AA'$ collisions as ~\cite{Aversa89,Aur00,pgNLO}:
\begin{eqnarray}
\label{hadX}
&E_{\pi}& \f{\dd \sigma_{\pi}^{AA'}}{ \dd ^3p} = 
            \int \dd ^2b \,\, \dd ^2r \,\, t_A(r) 
            \,\, t_{A'}(|{\bf b} - {\bf r}|) \,\, 
            \f{1}{s} \,\, \sum_{abc} \times \nonumber \\
& \times &  \int^{1-(1-v)/z_c}_{vw/z_c} \! \f{\dd \hat{v}}{\hat{v}(1-\hat{v})} \!
            \int^{1}_{vw/\hat{v}z_c} \f{ \dd \hat{w} }{\hat{w}} \!
            \int^1 {\dd z_c} \times \nonumber \\
& \times &  \int \!\! {\dd^2 {\bf k}_{Ta}} \!\! \int \!\! {\dd^2 {\bf k}_{Tb}}
            \,\, f_{a/A}(x_a,{\bf k}_{Ta},Q^2)
            \,\, f_{b/A'}(x_b,{\bf k}_{Tb},Q^2)  \nonumber \\
& \times &  \left[
            \f{\dd {\widehat \sigma}}{\dd \hat{v}} \delta (1-\hat{w})\, + \,
            \f{\alpha_s(Q_r)}{ \pi}  K_{ab,c}(\hat{s},\hat{v},\hat{w},Q,Q_r,\tilde{Q}) \right] \times \nonumber \\
& \times &
            \f{D_{c}^{\pi} (z_c, \tilde{Q}^2)}{\pi z_c^2}  \,\,  . 
\end{eqnarray}
Here $t_{A}(b) = \int \dd z \, \rho_{A}(b,z)$ is the nuclear thickness
function 
normalized as $\int \dd ^2b \, t_{A}(b) = A$.
For small nuclei we use a sharp sphere approximation, while for
larger nuclei the Wood-Saxon formula is applied.

In our next-to-leading order (NLO) calculation~\cite{pgNLO}, 
$\dd {\widehat \sigma}/ \dd \hat{v}$ represents the Born cross section
of the partonic subprocess and 
$K_{ab,c}(\hat{s},\hat{v},\hat{w},Q,Q_r,\tilde{Q})$ is the corresponding 
higher order correction term, see Ref.s~\cite{pgNLO,Aversa89,Aur00}. 
We fix the factorization and renormalization 
scales and connect them to the momentum of the intermediate jet, 
$Q=Q_r=(4/3) p_q$ (where $p_q=p_T/z_c$), reproducing $pp$ data
with high precision at high $p_T$ \cite{dAu}.

The approximate form of the 
3-dimensional parton distribution function (PDF) 
is the following:
\begin{equation}
f_{a/p}(x_a,{\bf k}_{Ta},Q^2) \,\,\,\, = \,\,\,\, f_{a/p}(x_a,Q^2) 
\cdot g_{a/p} ({\bf k}_{Ta}) \ .
\end{equation}
Here, the function $f_{a/p}(x_a,Q^2)$ represents the standard longitudinal NLO 
PDF as a function of momentum fraction of the incoming parton, $x_a$ at 
scale $Q$ (in the present calculations we use the
MRST(cg)\cite{MRST01} PDFs). The partonic 
transverse-momentum distribution in 2 dimensions, 
$g_{a/p}({\bf k}_T)$,  is characterized
by an "intrinsic $k_T$" parameter as in Ref.s~\cite{Yi02,pgNLO}. 
In our phenomenological approach this component 
is described by a Gaussian function~\cite{Yi02,Bp02}.

Nuclear multiscattering is accounted for through 
broadening of the incoming parton's transverse 
momentum distribution function, 
namely an increase in the width of the Gaussian:
\beq
\label{ktbroadpA}
\k2av_{pA} = \k2av_{pp} + C \cdot h_{pA}(b) \ .
\eeq
Here, $\k2av_{pp}=2.5$ GeV$^2$ is the width of the 
transverse momentum distribution
of partons in $pp$ collisions~\cite{Yi02,dAu}, $h_{pA}(b)$ 
describes the number of {\it effective} $NN$ collisions at impact 
parameter $b$, which impart an average transverse momentum squared $C$. 
The effectivity function $h_{pA}(b)$ can be written in terms of the 
number of collisions suffered by the incoming proton in the target nucleus.
In Ref.~\cite{Yi02} we have found  a limited number of semihard collisions,
$\nu_{m} = 4$ and the value $C = $ 0.4 GeV$^2$.

We take into account the isospin asymmetry by using a
linear combination of $p$ and $n$ PDFs. 
The applied PDFs are also modified inside nuclei
by the ``shadowing'' effect\cite{Shadxnw_uj}. 

The last term in the convolution of eq. (\ref{hadX}) is
the fragmentation function (FF), $D_{c}^{\pi}(z_c, \tilde{Q}^2)$. This gives 
the probability for parton $c$ to fragment into a pion with momentum 
fraction $z_c$ at fragmentation scale $\tilde{Q}=(4/3) p_T$. We apply the
KKP parametrization~\cite{KKP}.

\section{Jet-Quenching as a Final State Effect}
\label{sec:2}

The energy loss of high-energy quark and gluon jets traveling through  
dense colored matter is able to give information on the density of 
gluons~\cite{tomog}. This 
non-Abelian radiative energy loss $\Delta E (E,L) $ can be described as 
a function of gluon density:  $\bar{n}=L/ \lambda_g $, the mean number of jet 
scatterings, where $L$ is the length 
traversed by the jet and $\lambda_g$ is the mean 
free path in non-Abelian dense matter. In "thin plasma" approximation
energy loss in first order is given by the following form~\cite{glv}:
\begin{eqnarray}
\label{glv}
\Delta E^{(1)}_{GLV}
&=& \frac{2 C_R \alpha_s}{ \pi} \frac{EL}{\lambda_g} \int\limits^1_0 \dd x 
    \int\limits^{k_{max}^2}_0 \frac{\dd {\bf k}^2_T}{{\bf k}^2_T} \times 
    \nonumber \\ 
& \times & \int\limits^{q^2_{max}}_0 \frac{\dd ^2 {\bf q}_T \mu^2_{eff}}
    { \pi \left( {\bf q}^2_{T} + \mu^2 \right) ^2} \cdot
    \frac{2 {\bf k}_T \cdot {\bf q}_T \left( {\bf k} -{\bf q} \right) ^2_T L^2}
    {16\,x^2\,E^2 +\left( {\bf k} -{\bf q} \right) ^4_T L^2} \nonumber \\  
&=& \frac{C_R \alpha_s}{N(E)} \frac{L^2 \mu^2}{\lambda_g} 
\log\left( \frac{E}{\mu} \right) \,\,\, ,
\end{eqnarray}
where $C_R$ is the color Casimir of the jet, 
$\mu/\lambda_g \sim \alpha_s^2 \rho_{part}$ 
is a transport coefficient of the medium, 
proportional to the parton density, $\rho_{part}$. 
The color Debye screening scale is $\mu$,
and $\lambda_g$ is the radiated gluon mean free path. 
$N(E)$ is an energy dependent factor with asymptotic value 4.   

Considering a time-averaged, static plasma, 
the average energy loss, $\Delta E$,
will modify the argument of the FFs:
\begin{equation}
\label{quenchff}    
\frac{D_{\pi/c} ( z_c , \tilde{Q}^2 )}{\pi z_c^2 } \longrightarrow 
\frac{z^{\ast}_c}{z_c}  \,\, 
\frac{D_{\pi/c} ( z^{\ast}_c , \tilde{Q}^2 )}{\pi z_c^2 }.
\end{equation}
Here $ z^{\ast}_c = z_c / \left(1- \Delta E/p_c \right) $
is the modified momentum fraction.

In Fig. 1 we present our result on pion production  
in most central ($0-10\%$) $AuAu$ collisions at $\sqrt s =$ 200 AGeV. 
We included all initial and final state effects discussed above
and used the opacity value $\bar{n}=3.5 \pm 0.25$ to reproduce the
experimental data.
\begin{figure}
\resizebox{0.40\textwidth}{!}{%
  \includegraphics{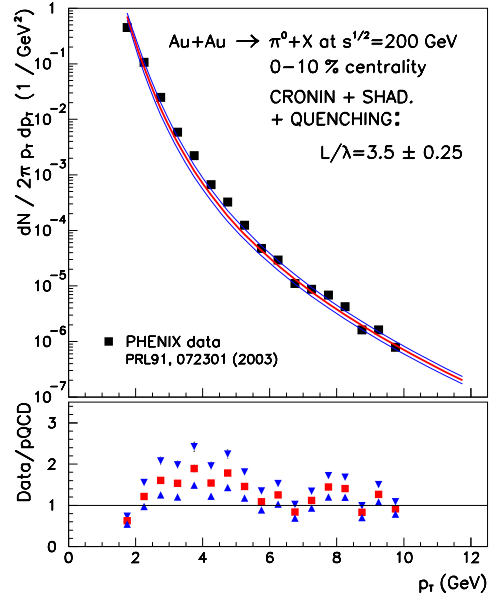}
}
\caption{Pion production in central $AuAu$ collision 
with the calculated opacities ${\bar n}=3.5 \pm 0.25$
(upper panel). Data are from PHENIX Collaboration~\cite{phenix0203}. 
The lower panel displays a comparison between the data and the
calculations.}
\label{fig:1}       
\end{figure}

\begin{figure}
\resizebox{0.48\textwidth}{!}{%
  \includegraphics{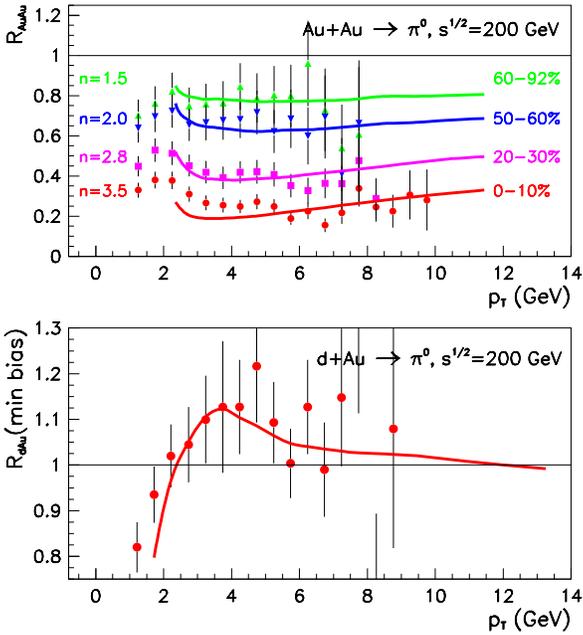}
}
\caption{Upper panel displays the 
pion production in $AuAu$ collision in different centrality bins
with the calculated opacities.
Data are from PHENIX Collaboration~\cite{phenix0203}. 
Lower panel shows the pion production in dAu collision~\cite{phenixdau} and
the result of our calculation~\cite{dAu}.}
\label{fig:2}       
\end{figure}

\section{Centrality Dependence in $AuAu$ collisions}
\label{sec:3}

In the top panel of Fig. \ref{fig:2} we display
the nuclear modification factor 
\begin{equation}
R_{AA'}(p_T,b)=\frac{1}{N_{bin}} \cdot 
\frac{E_{\pi} \dd \sigma^{AA'}_{\pi}(b)/\dd^3 p_T}
{E_{\pi} \dd \sigma^{pp}_{\pi}/\dd^3 p_T}, 
\end{equation}
(where $N_{bin}$ is the number of binary collisions),
as a function of $p_T$ at different impact parameter ranges.
The measured suppression is reproduced in the most central collisions 
with opacity  ${\bar n}=3.5 \pm 0.25$ as it was shown in Fig. \ref{fig:1}.
Curves with lower values of opacities are shown for comparison
to the more peripheral data. 
Since the centrality $60-92\%$ includes non-peripheral events, the
obtained  ${\bar n}=1.5$ seems to be a reasonable 
opacity value for this bin. In the very peripheral case the opacity
should be reduced to a small value~\cite{bgg}, 
however  $\pm 20$\% error on the
nuclear modification factor does not allow more detailed investigation.
A more quantitative analysis can be performed, 
when data will be available with smaller error bars.
At higher precision, the geometry of the
hot overlap zone can be included and a more detailed
analysis of the impact parameter dependence can be accomplished.
In this case the properties of the produced hot matter will be studied
in a more quantitative way.




In the bottom panel of Fig. \ref{fig:2} the $dAu$ data  
are compared to our calculation~\cite{dAu}
to demonstrate the validity
of our description for the initial nuclear effects. 

\section*{Acknowledgments}

First of all, let us thank the Organizers and EPS 
for this informative meeting and the support of 
the participation of one of the authors (GGB). 
This work was supported in part by Hungarian 
grants T034842, T043455, T043514, U.S. DOE grant: DE-FG02-86ER40251, 
and NSF grant: INT0000211. Supercomputer time provided by BCPL in 
Norway and the EC -- Access to Research Infrastructure action of the 
Improving Human Potential Programme is gratefully acknowledged.

%

\end{document}